\documentclass[a4paper, 10pt, conference]{IEEEtran}
\ifCLASSINFOpdf
\usepackage[pdftex]{graphicx}
\graphicspath{{images/}}
\DeclareGraphicsExtensions{.pdf,.jpeg,.png}
\else
\usepackage[dvips]{graphicx}
\graphicspath{{eps/}}
\DeclareGraphicsExtensions{.eps}
\fi

\usepackage{balance}
\usepackage{stfloats}
\usepackage{float}
\usepackage{url}
\usepackage{enumerate}
\usepackage{setspace}
\usepackage{balance}
\usepackage{caption}
\usepackage{xcolor}
\usepackage{graphicx}
\usepackage{algpseudocode}
\usepackage{color}
\usepackage{multicol}
\usepackage{multirow}
\usepackage[linesnumbered,ruled,vlined]{algorithm2e}
\usepackage[flushleft]{threeparttable}

\usepackage{setspace}
\usepackage{balance}
\usepackage[utf8]{inputenc}

\usepackage[scaled]{beramono}
\usepackage[T1]{fontenc}
\usepackage{setspace}
\usepackage{amsmath}
\usepackage{listings}
\usepackage{textcomp}
\usepackage[utf8]{inputenc}
\usepackage{tabulary}
\usepackage{enumitem}
\usepackage{tikz}
\newcommand*\circled[1]{\tikz[baseline=(char.base)]{
		\node[shape=circle,draw,inner sep=1pt] (char) {#1};}}
	
\setlist{nolistsep}
\def\baselinestretch{0.98}

\usepackage{draftwatermark}
\SetWatermarkScale{1}
\SetWatermarkColor[gray]{0.9}

\newif\ifconf
\conftrue

\begin{document}

\title{ CloudCAMP: Automating Cloud Services Deployment \& Management} 


\author{
	\IEEEauthorblockN{Anirban Bhattacharjee\IEEEauthorrefmark{1}, Yogesh Barve\IEEEauthorrefmark{1}, Aniruddha Gokhale\IEEEauthorrefmark{1}}
	\IEEEauthorblockA{\IEEEauthorrefmark{1}Department of Electrical Engineering and Computer Science\\
		Vanderbilt University, Nashville, Tennessee, USA\\
		Email: {\{anirban.bhattacharjee; yogesh.d.barve; a.gokhale\}}@vanderbilt.edu
	}
	\and 
	\IEEEauthorblockN{Takayuki Kuroda\IEEEauthorrefmark{2}}
	\IEEEauthorblockA{\IEEEauthorrefmark{2}NEC Corporation \\
		Kawasaki, Kanagawa, Japan\\
		Email: t-kuroda@ax.jp.nec.com}
}

\maketitle

\begin{abstract}

Users of cloud platforms often must expend significant manual efforts in the deployment and orchestration of their services on cloud platforms due primarily to having to deal with the high variabilities in the configuration options for virtualized environment setup and meeting the software dependencies for each service.  Despite the emergence of many DevOps cloud automation and orchestration tools, users must still rely on specifying low-level scripting details for service deployment and management using Infrastructure-as-Code (IAC).  Using these tools required domain expertise along with a steep learning curve.   To address these challenges in a tool-and-technology agnostic manner, which helps promote interoperability and portability of services hosted across cloud platforms, we present initial ideas on a GUI based cloud automation and orchestration framework called CloudCAMP.  It incorporates domain-specific modeling so that the specifications and dependencies imposed by the cloud platform and application architecture can be specified at an intuitive, higher level of abstraction without the need for domain expertise using Model-Driven Engineering(MDE) paradigm. CloudCAMP transforms the partial specifications into deployable Infrastructure-as-Code (IAC) using the Transformational-Generative paradigm and by leveraging an extensible and reusable knowledge base.  The auto-generated  IAC can be handled by existing tools to provision the services components automatically. We validate our approach quantitatively by showing a comparative study of savings in manual and scripting efforts versus using CloudCAMP.  

\end{abstract}

\renewcommand\IEEEkeywordsname{Keywords}



\section{INTRODUCTION}
\label{sec:CloudCAMPintroduction}

Self-service application deployment and management in a fault-free manner is desired for enterprises to speed up time-to-market for their cloud services. Enterprises often suffer from service outages and delays that stem predominantly from the use of tedious and error-prone manual efforts that are expended in service configuration, integration, and infrastructure provisioning across the heterogeneous platforms.  Modern cloud services are architected as microservices, and each of the components must be configured and deployed on cloud platforms -- sometimes federated -- in a specific order. The capabilities of the entire service are realized through a collection of distributed, loosely coupled service components~\cite{barve2017pads,bhattacharjeemde}. Script-centric efforts to deploy and manage these complex scenarios degrade productivity and adversely impact the product time-to-market.  


\subsection{Motivating the Problem}
\label{sec:CloudCAMPmotivation}

\begin{figure}[htb]
	\centering
	\includegraphics[width=0.9\linewidth]{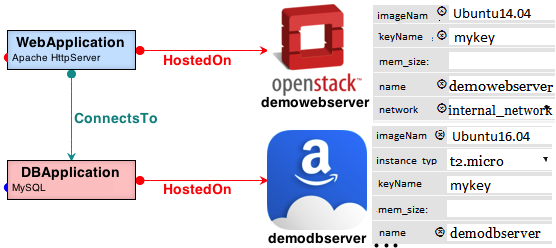}
	\caption{Desired Level of Abstraction for a WebApp Business Model}
	\label{Fig:business_model}
\end{figure}

Consider the case of a LAMP [Linux, Apache, MySQL, and PHP] -based service deployment on a cloud
platform. Figure~\ref{Fig:business_model} shows the desired cloud application topology consisting of two connected software stacks, i.e., a PHP-based web
front-end and a MySQL database backend.  The frontend WebApplication stack holds the business logic, and it will be deployed on Ubuntu 16.04 server virtual machine (VM), which is managed using the OpenStack cloud platform. The backend DBApplication stack holds the relational database, which is used to store and query the product data.  The backend database is a MySQL DBMS, which
will be deployed on the Amazon Elastic Compute Cloud (EC2) VM instance with an Ubuntu 14.04 server. 

\subsection{Requirements for CloudCAMP}
\label{sec:CloudCAMPchallenges}

Based on this use case, we elicit the following requirements that drive the
solution presented in this paper. 

\subsubsection{Requirement 1: Reduction in specification details needed for deployment} \label{sec:req1}

\begin{figure}[htb]
	\centering
	\includegraphics[width=0.9\columnwidth]{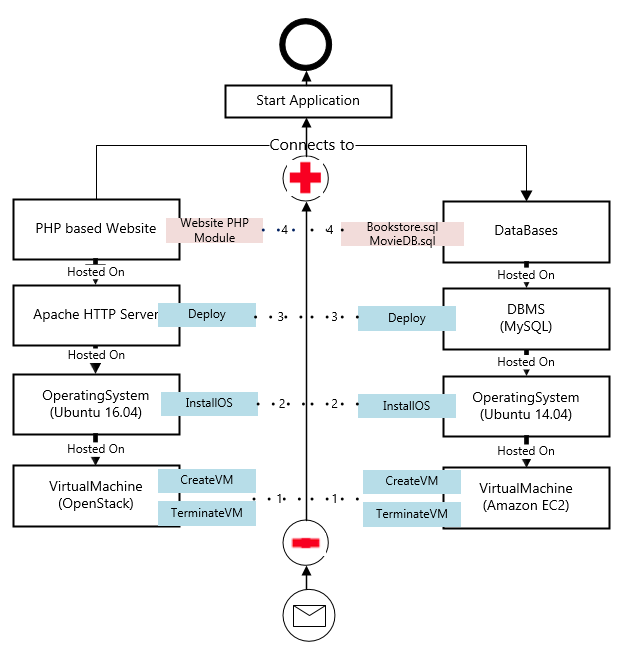}
	\caption{A TOSCA-compliant PHP- and MySQL-based Application
		Deployment Workflow} 
	\label{Fig:motivation}
\end{figure}

As depicted in Figure~\ref{Fig:motivation}, the deployer needs to provision the PHP and MySQL-based e-commerce application stack from two aspects. In the cloud infrastructure provisioning aspects, the
application topology needs to be woven into the execution environment which can be virtual machines (VM), containers or third party services.  To provision the cloud infrastructure, the deployer needs to select a proper image for their VM, along with the security group, roles, network, number of instances, the storage unit in the target cloud providers' platform.  In the service provisioning aspects, all the dependent software needs to be
installed, and all the constraints need to be configured.  For example, for the
frontend of our motivating example, Apache Httpd needs to be installed and
configured along with PHP and Java. Similarly, in the backend, MySQL needs to
be installed and configured. Moreover, the database service should start before the PHP
application service so as to run the WebApp properly. The IAC solution for the
application provisioning requires all the installation and configuration
details to execute the deployment plan.

This scenario shows that a
user must possess extensive domain knowledge to provision even a simple web
application correctly, and we aim to abstract these detailed specifications
from the users.

\subsubsection{Requirement 2: Auto-completion of Infrastructure Provisioning } \label{sec:req2}
Writing the low-level intricate scripts to provision the infrastructure for the
motivating scenario is time-consuming.  To improve productivity by
significantly alleviating such efforts requires a self-service framework, which
should be capable of transforming the abstracted business models to complete,
deployable TOSCA-compliant\footnote{TOSCA~\cite{toscaspec} is an OASIS standard
  for vendor-neutral topology and orchestration specification for cloud-based
  applications.} Infrastructure-as-Code (IAC)
solution~\cite{breitenbucher2013integrated}. 

\subsubsection{Requirement 3: Support for Continuous Integration, Migration, and Delivery}\label{sec:req3}
Suppose that for the use case of Figure~\ref{Fig:motivation}, the enterprise
wants to execute a management task to migrate the web front-end to Amazon's EC2
with the purpose of reducing the number of cloud providers used by their
services.  To migrate the frontend, the user must perform the following steps:
(1) shut down the old virtual machine on OpenStack,
(2) create a new virtual machine on Amazon EC2, 
(3) install the Apache HTTP server and the other dependencies, 
(4) deploy the PHP-based frontend, and so on.

This migration activity gives rise to several issues, such as having to deal
with missing database drivers and missing configurations of the target database
service.   
Manually performing the migration tasks often require sheer technical expertise
about the different cloud APIs and its underlying technologies.  Application
extensibility (such as adding one database server node or data analysis
toolkits with the existing application) will also incur additional challenges.

All of these challenges motivate the need for a fully automated platform that can generate the robust deployment plans.
Nevertheless, the challenge here also lies in capturing the
application and cloud specifications in the metamodel and the
DSML. However, apart from that, there are a few more problems, which
are listed below: 

\paragraph {\textbf{Extensibility and Reusability of the Application Components}} \label{sec:req2a}
New application components need to be added at runtime to the existing
application by leveraging the platform. The specifications captured in
the metamodel should be modularized and loosely coupled with a
particular application. DSMLs should do all the binding after querying
for the specification for particular application type in the knowledge
base, and then the DSML will generate concrete cloud-specific,
operating-system specific infrastructure-as-a-code solution. The IAC
is idempotent, so it will not change the existing deployment if
configured correctly. The correctness of the added application
components can be validated using constraint checker at the model
level. 

\paragraph{\textbf{Extensibility of the Platform} }  \label{sec:req2b}
The platform can transform the business-relevant model to actionable
infrastructure-as-a-code, which produces application deployments in
the cloud.  However, the challenge is to make the platform loosely
coupled with any DevOps or orchestrating tool, so that later different
tools can be added if required.  Moreover, adding new application
requires reverse engineering the application components, and capturing
the application specifications in the metamodel of our platform, and
adding new cloud providers also requires a similar approach. Defining
commonality and variability points is critical to building a
modularized platform so that extensibility of platform will be
relatively easy. 

In our proof-of-concept solution, we only generate the Ansible
specific code from the business model, and our WebGME metamodel
handles the TOSCA specification.

\begin{figure*}[htb]
	\centering
	\includegraphics[width=0.9\textwidth,keepaspectratio]{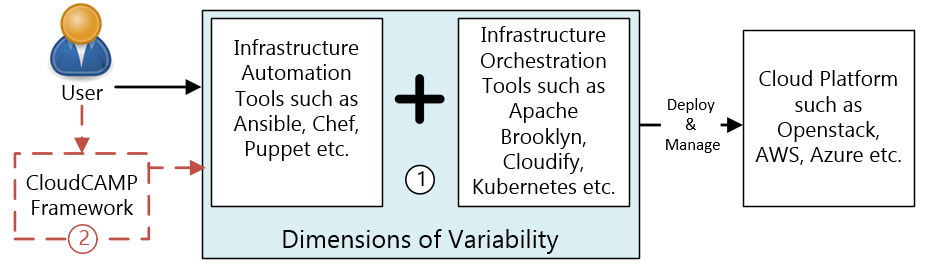}
	\caption{\color{blue}{Box 1} \color{black}depicts the responsibilities of service deployment team, which is to define the low-level scripts so that existing automation tools can configure the application components and orchestration tools can provision the infrastructure for application components and execute them on heterogeneous cloud environments.  \color{red}{Box 2} \color{black} depicts the contributions of this paper which introduces a self-service  framework and automates whole infrastructure design solutions for these tools.}   
	\label{Fig:selfservice}
\end{figure*}

\subsection{Limitations of Existing Approaches}
Self-service application provisioning requires extensive planning for their smooth operations.  In the context of cloud-based service hosting, service provisioning includes two key steps: (a) orchestration, where the deployment and ordering of individual components of the service must be managed across distributed resources of a cloud platform or federated cloud platforms, and (b) service automation, where defining and executing individual resource-specific configurations, such as a virtual machine or container configurations, and deployment of service components on these resources are automated. \emph{Infrastructure as Code (IAC)} is a term used by the DevOps community in which the cloud infrastructure is viewed as software for which code is developed to automate the entire cloud-based service provisioning.  To that end the DevOps community today leverages orchestration solutions such as Cloudify, Apache Brooklyn, and Kubernetes, among others in conjunction with automation tools such as Ansible, Puppet, and Chef, among others.  This state of the art (i.e., the extensive choices available to the developer) is reflected in Figure~\ref{Fig:selfservice}.   

The choice of services provided by different cloud providers needs to be
selected and configured by the deployer. It requires elaborate specifications
of service topologies comprising requirements, functionalities, dependencies
and relationships of the components. For instance, depending on the technology
used, e.g., MySQL versus PostgreSQL or PHP versus Node.js, the script must
include the appropriate drivers. Also, architecting the solution for different
cloud providers is different.  For instance, creating data flow architecture
using AWS Kinesis and DynamoDB is much different from creating the same
architecture using Azure Event Hubs and CosmosDB. Additional dimensions of
variability (i.e., addressing application's compatibility and cloud providers'
incompatible APIs) as depicted in Box 1 of Figure~\ref{Fig:selfservice}
complicates the manual effort which is already daunting, tedious and
error-prone.  Finally, existing approaches do not account for pre-deployment
validation to check if the end-user requirements and software dependencies are
met.    

\subsection{Solution Approach: CloudCAMP}
Our motivating example shows that a user must possess extensive domain knowledge to provision even a simple web application stack correctly. Users need to write Infrastructure-as-Code (IAC) solutions via low-level scripting.   Instead, the desired capability would require a self-service platform, in which (1) a deployer specify only the application components, such as a Web App, and (2) the framework automatically transforms the business model into deployable artifacts. To achieve the goal, we propose a model-driven and scalable, rapid provisioning framework called CloudCAMP.
It complies with TOSCA (Topology and Orchestration Specification for Cloud Applications) specification, which enables the creation of portable and interoperable \emph{plans-as-a-service} template for cloud services. TOSCA provides the standardization for decoupling software applications and its dependencies from the cloud platform specifications.   

The key contributions in this paper include:
\begin{enumerate}
	\item We present key elements of CloudCAMP's domain-specific modeling
	language (DSML) that masks low-level details of the application component
	specifications and cloud provider specifications and instead offers
	intuitive high-level  representations; 
	\item We present the use of an extensible knowledge base and algorithms to
	perform Model-to-Infrastructure-as-code (IAC) transformations
	automatically; and 
	\item We present a concrete realization of CloudCAMP and validation in the
	context of real-world use cases. 
\end{enumerate}

\subsection{Organization of the paper}
The rest of the paper is organized as follows:    
Section~\ref{sec:CloudCAMPsurvey} presents a brief survey of existing literature and compares to our solution;     Section~\ref{sec:CloudCAMParch} presents the design of CloudCAMP;   
Section~\ref{sec:CloudCAMPcasestudies} evaluates our metamodel for a prototypical case study and presents a user survey;    and finally, Section~\ref{sec:CloudCAMPconclusion} concludes the paper alluding to future
directions.


\section{RELATED WORK}
\label{sec:CloudCAMPsurvey}

The problem of deployment and management abstraction has been explored in the
area of cloud automation and orchestration. In this section, we compare
existing efforts in the literature with our work.   
The use of these toolchains adds the burden of configuring the application components and integrating
pre-deployment verification on application developers.    
The script-centric DevOps community provides toolchains for eliminating the
disconnect between developers and operations
providers~\cite{humble2011enterprises}, these tools incur limitations in
providing a self-service provisioning platform. Cloud orchestration tools like
 Apache Scalr
(\url{https://scalr-wiki.atlassian.net/wiki/display/docs/Apache}),  
CloudFoundry (\url{https://www.cloudfoundry.org/}), Cloudify
(\url{http://getcloudify.org/}) etc. are excellent toolchains to deploy and
manage applications on any cloud providers.   They provide techniques to
monitor the health of the application and to migrate between the cloud
providers using standardized approaches. However, they all suffer from the
limitations of requiring the users to define the complete and correct
deployable model with all the functionalities and features. In this context,
Alien4Cloud~\cite{carrasco2016bidimensional} proposes a visual way to generate
TOSCA topology model, which can be orchestrated by Apache Brooklyn.  However,
building the proper topology even using an MDE approach
combined with the TOSCA specification  
needs domain expertise.  Unlike these approaches, CloudCAMP abstracts all the
application and cloud-specific details in the metamodel of its DSML and
transforms the business model to TOSCA-compliant IAC.

Several patterns-based approaches are proposed to reduce the complexity of service deployment~\cite{eilam2011pattern, lu2013pattern}. They differentiate between business logic and the deployment of service-oriented architecture
platform. 
Each pattern offers a set of capabilities, and characteristics.  
Likewise, model-based patterns of proven solutions
are used for service deployment in cloud infrastructures~\cite{kepes2017sepade,leite2014deploying}.  For instance, MODAClouds~\cite{ardagna2012modaclouds} allows users to design, develop and re-design application components to operate and manage in multi-cloud environments using a Decision Support System.   
In Computation Independent Model, the design artifacts are
semi-automatically translated to Cloud-Provider Independent Model
level, where an entirely deployable abstract cloud model is generated
by matching the application patterns.  The abstract deployment model
is concretized to Cloud-Provider Specific Model (CPSM) by a
domain-specific language.   
Similar to CloudCAMP, they also support reuse and role-based iterative refinement in a component-based approach. However, their deployment plan generation lacks verification and extensibility.  They also did not consider distributing application components in a heterogeneous cloud environment.

Several efforts come close to the CloudCAMP idea.  For instance,
ConfigAssure~\cite{narain2008declarative} is a requirement solver to synthesize infrastructure configuration in a declarative fashion.  All the requirements are expressed as constraints by the developer, and the provider predefines a configuration database containing variables as a deployment model. Kodkod~\cite{torlak2007kodkod} is a relational model finder which takes these arguments as a first-order logic constraint in the finite domain. 
Engage~\cite{fischer2012engage} deploys and manages the application from a partial specification using a constraint-based algorithm.   
Aeolus Blender~\cite{di2015automatic} comprises the configuration optimizer Zephyrus~\cite{di2014automated}, the ad-hoc planner Metis~\cite{lascu2013planning}, and deployment engine Arnomic.  Zephyrus automatically generates an abstract configuration of the desired system based on a partial description. 
They guarantee meeting all the end-user requirements for software dependencies and provide an optimal solution for a given number of active virtual machines.  
In contrast to the use of the knowledge base in CloudCAMP, these efforts use a CSP solver to transform the business model.  CSP solvers, however, can take significant time to execute.  Moreover, defining constraints on the configurations requires domain expertise, which is not needed in CloudCAMP.    

Similar to CloudCAMP, Hirmer et al.~\cite{hirmer2014automatic} focus on producing complete TOSCA-compliant topology from users' partial business relevant topology. Users have to specify the requirements directly using definitions of the corresponding node types or are added manually for refinement.  
Their completion engine compares user specification with target models and combines the missing components to make it a fully deployable model, and then the service components can be executed in the right order using an OpenTOSCA toolchain~\cite{breitenbucher2014combining}. CELAR~\cite{giannakopoulos2014celar} combines MDE and TOSCA specification to automate deployment cloud applications, where topology completion is fulfilled by requirement and capability
analysis on node template.  Unlike these efforts, the model transformation in CloudCAMP is based on querying the knowledge base and idempotent infrastructure code generation.


\section{CloudCAMP DESIGN and IMPLEMENTATION}
\label{sec:CloudCAMParch}
\begin{figure*}[!htb]
	\centering
	\includegraphics[width=\textwidth]{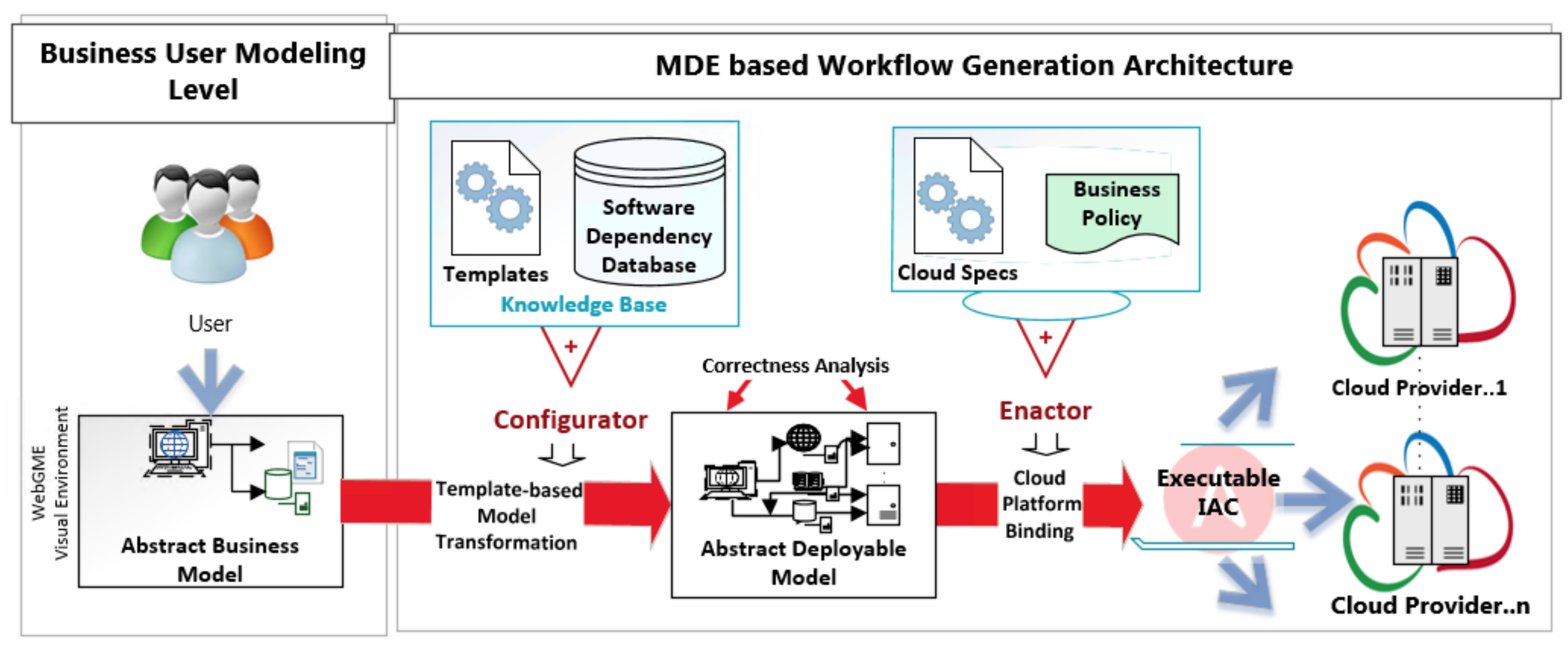}
	\caption{The CloudCAMP Workflow}
	\label{Fig:plan}
\end{figure*}

This section delves into the design details of CloudCAMP (Figure~\ref{Fig:plan}) and shows how it meets the requirements discussed in Section~\ref{sec:CloudCAMPchallenges}.

\subsection{System Architecture} 
\label{sec:workflow}  
To better appreciate the CloudCAMP solution presented below, consider the fundamental requirements outlined earlier and depicted in Figure. As per our framework design, 1) A deployer needs to specify only the application components, such as a Web App, using intuitive notations provided by the framework, 2) The DSML transforms the business model into deployable artifacts. Thus, the first step is for the user to utilize an intuitive, higher-level modeling framework that simplifies the modeling of business logic and automatically takes care of non-business centric deployment and management artifacts.  

To that end, we have architected CloudCAMP's cloud-based service provisioning workflow as depicted in Figure~\ref{Fig:plan}.  Below we explain the roles of the different actors involved~\cite{bhattacharjee2018wip}:

\renewcommand{\labelenumii}{\Roman}
\begin{enumerate}[label=\protect\circled{\arabic*}]
    \item \emph{Business User Modeling}: A business application is modeled as a compendium of different application components. The user has to select appropriate application component types from the CloudCAMP application pane to deploy the associated application code.  The user needs to specify the variability points for application components' deployment as depicted in Figure~\ref{Fig:business_model}. The design details of CloudCAMP DSML is described in Section~\ref{sec:meta1}.
	
    \item \emph{Configurator}: This actor is responsible for transforming each abstract description of an application component to a deployable cloud automation task (e.g., Ansible-specific) for each application component. Configurator realizes a user-defined abstract description of a cloud application model, and then maps the application components with the operating system, and query the knowledge base to find the software dependency tree; it generates full `correct-by-construction' Ansible specific code from the application type template. The details of template-based transformation and code generation are described in Section~\ref{sec:meta2}.
	
    \item \emph{Enactor}: It generates the infrastructure design workflow of IAC solutions by integrating the generated automation code with the business rules and cloud infrastructure specifications. The users define the connection types between the application components. There are four types of connections: `hostedOn', `connectsTo', `deleteFrom', `migrateTo'. The details of the connection types and their role is described in Section~\ref{sec:relations}. The orchestration tool executes all the automation tasks based on the connection types to deploy and run the business application components in proper order.  
	
    \item \emph{Knowledge Base\label{kb}:} A knowledge base is needed for auto-completing the partially specified deployment models.  We predefine the software dependencies for application type in a relational table with a key-value pair.  All the software packages needed for a particular application component are defined in the tables along with their dependency on the operating system and its version.  The application developer needs to populate the tables with all software dependencies for including the new application component type in the CloudCAMP. The design details of the knowledge base are described in Section~\ref{sec:kb}.
	
\end{enumerate}

\begin{figure*}[htb]
	\centering
	\includegraphics[width=0.9\textwidth]{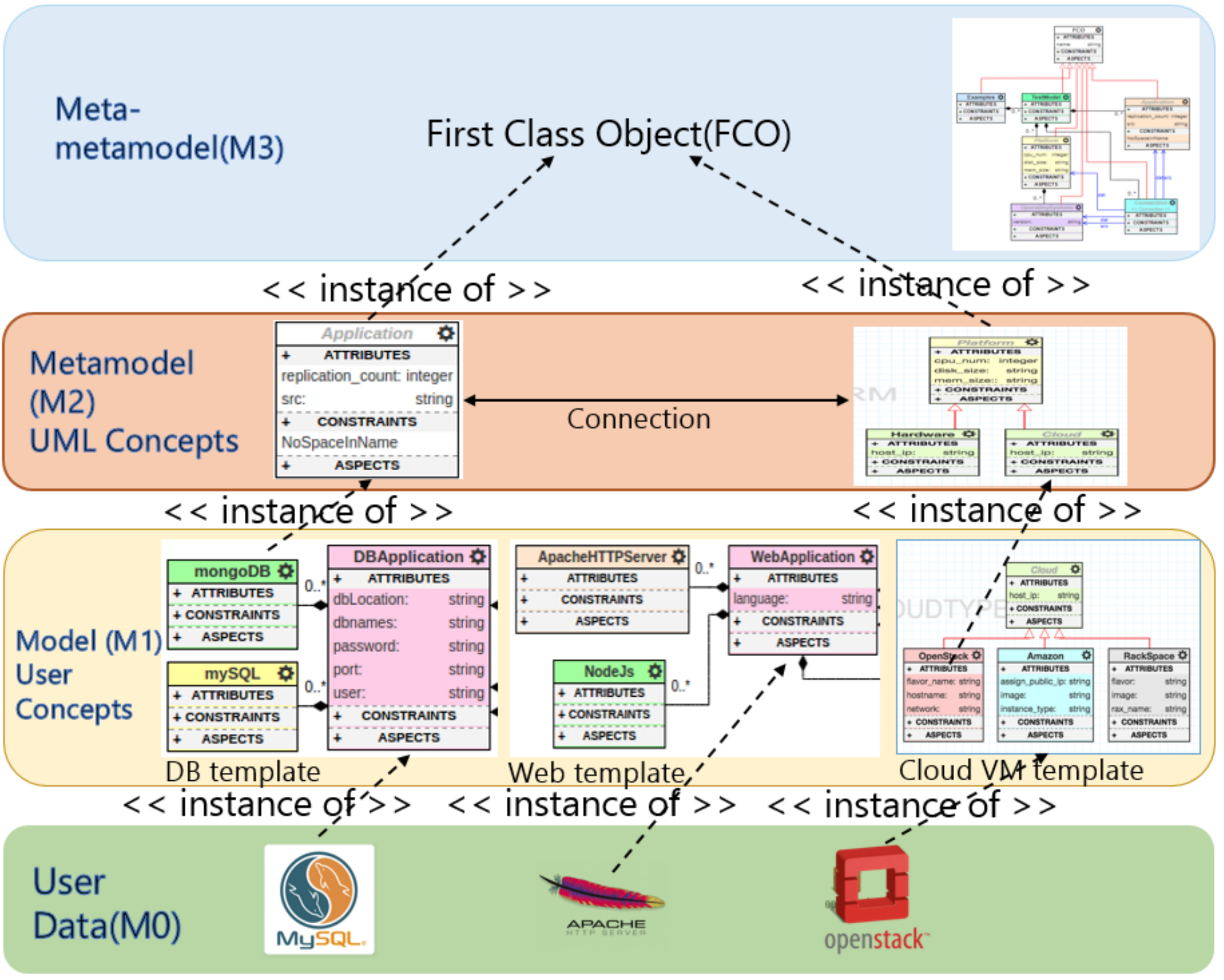}
	\caption{A Partial Meta-Object Facility (MOF) model of CloudCAMP DSML and Platform}
	\label{Fig:mof}
\end{figure*}

\subsection{System Implementation} 
The CloudCAMP DSML shown in Figure~\ref{Fig:mof} is developed using the WebGME MDE framework (\url{www.webgme.org}). WebGME is a cloud-based framework that offers an environment for DSML developers to define their language and create model parsers that can serve as generators of code artifacts.  The CloudCAMP runtime platform uses a microservices architecture comprising three services: (a) the modeling infrastructure, i.e., the WebGME UI, and orchestration and automation frameworks forming one service, (b) the WebGME modeling details are stored in a MongoDB NoSQL database, and (c) the knowledge base is hosted as a MySQL database service.  The microservices are connected through the API endpoints and placed behind a HAproxy (\url{http://www.haproxy.org/}) load balancer.  Thus, all the services can independently scale to support parallel spawning and configuration of multiple VMs or containers in the cloud platform.

\subsection{CloudCAMP Domain-specific Modeling Language (DSML)} 
\label{sec:meta1}  

The CloudCAMP DSML abstracts the design complexities by separating the application from deployment and infrastructure technologies according to TOSCA specification as described in Requirement~\ref{sec:req1}.

\subsubsection{Design Rationale for CloudCAMP Metamodels }
DSMLs are realized through one or more interrelated metamodels that capture the DSML's syntax and semantics.  In our case, to transform the business model to a full-blown deployment model, we needed to capture various facets of the application and cloud specifications in our metamodel.  CloudCAMP's deployment modeling automation metamodel was developed by harnessing a combination of (1) reverse engineering, (2) dependency mapping across heterogeneous clouds, (3) dependency mapping across different operating systems and their versions, (4) semantic mapping, (5) business policy, and (6) prototyping.  Capturing this variability helps to enrich the expressive power, multi-cloud tool support and interoperability of the platform.  Prototyping and reverse engineering helped to identify the different application components, cloud and operating system specific endpoints.  The dependent software packages, their relationship mapping, and configuration templates were realized in the metamodel by querying the knowledge base.  The set of available building blocks, requirements, policies, and other information concerning the implementation of the services and all other known constraints are pre-defined in the high-level application metamodel.   

To that end, CloudCAMP provides different node types, which are the application components such as Web Application, Database Application, DataAnalytics Application, and various cloud providers such as OpenStack, Amazon AWS, Microsoft Azure.  The goal is to concretize the abstract application node type by matching the application deployers' desired specification with the pre-defined functionalities captured in the CloudCAMP metamodel and knowledge base.  The concrete node templates are then woven to specific cloud provider types, and their VMs to create a dependency graph that has to be executed to deploy the application on the desired target machine as shown in Figure~\ref{Fig:plan}. Using our DSML, the deployer can configure the node in a defined cloud platform or particular target system with ease.

Snippets of the metamodels for CloudCAMP are shown in the M1 and M2 level of Figure~\ref{Fig:mof}, which are based on the Meta-Object Facility (MOF) standard provided by Object Management Group (OMG [\url{ http://www.omg.org/}]).  Using our DSML, the application deployer can configure the node in a defined cloud platform or particular target system without providing any deployment or implementation artifacts that contain code or logic.

\begin{figure}[htb]
	\centering
	\includegraphics[width=0.8\linewidth]{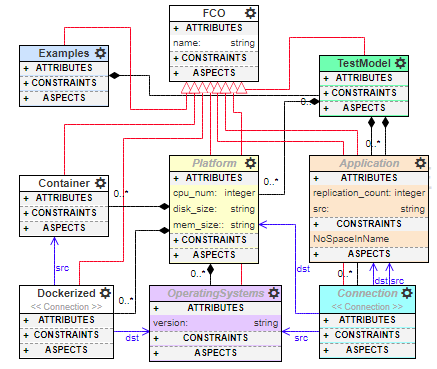}
	\caption{Main MetaModel of CloudCAMP framework. The black lines depict containment, the {\color{red}red} lines depict inheritence and {\color{blue} blue} lines depict connection.}
	\label{Fig:metamodel}
\end{figure}

The high-level metamodel is depicted in the
Figure~\ref{Fig:metamodel}. It shows the M2 and M3 level of the MOF
standard. The First Class Object(FCO) is the root node, and under it
the metamodel for the cloud platform, operating systems,  containers and
application components (M2 level) are defined. The connection type is
also defined at the M2 level. We will now dive down into the metamodel
for Cloud Platforms and Application components.   

The CloudCAMP metamodels are extensible and reusable, so new component types and platforms can be added as required in the CloudCAMP metamodel.

\subsubsection{Metamodel for the Cloud Platforms }
In designing the metamodel for cloud platforms, we observed (i.e., reverse
engineered) the process of hosting applications across different cloud
environments, and captured all the commonalities and variabilities.  The
specifications for different cloud platforms such as OpenStack, Amazon AWS,
Microsoft Azure, etc for provisioning virtual machines (VMs) with different
operating systems (OS) are captured. The deployers can choose their desired OS
images to spawn the VMs/containers. 

{\small
	\begin{verbatim}
	_cloudSpec_(type, vmtype, services, ostype)
	type ::= Openstack | Amazon | Azure | Hardware
	services ::= Function
	vmtype ::= Function
	os_type ::= ubuntu | redhat | windows
	\end{verbatim}	}

The deployer can select a pre-defined VM flavor, available networks, security
groups, roles, and the available images, all of which are defined as
variabilities in our metamodel.  They also must specify their environment file,
the secret key for the selected cloud host types, which are the endpoints to
bind to a particular cloud provider as shown in
Figure~\ref{Fig:business_model}. Optionally, a pre-deployed machine can be
specified by providing the IP address and OS. Available services and VM types
for cloud platforms are pre-defined in the metamodel. 

\subsubsection{Metamodel for Application Components }

For cloud-hosted services, CloudCAMP provides different node types for
application components such as Web Application, Database Application,
DataAnalytics Application, etc.  For instance, the metamodel enables a deployer
to choose the web server attribute, language for the code, the database server
attribute or the NoSQL database attributes from the provided list.  The
deployer has to specify the variable attributes to deploy the desired
application component type.

{\small
	\begin{verbatim}
	_appSpec_(type, os_type, swdependency, attributes)
	type ::= Web | Database | DataAnalytics | ...
	os_type ::= ubuntu | redhat | windows
	swdependency ::= Query to knowledgeBase
	attributes ::= Function
	\end{verbatim}	}

\subsubsection{Defining the Relationship among Components}\label{sec:relations}
Four relationship types bind the node types in the metamodel as follows:
\begin{enumerate}
	\item \emph{`hostedOn'} relationship type implies the source node type is required to be deployed on the destination node type, e.g.,  Webserver is hosted on Ubuntu 16.04 in OpenStack.    
	
	\item \emph{`connectsTo'} relationship type is used for deployment ordering to relate the source node type's endpoint to the required target node type endpoint if they are dependent on each other.   The node types linked by `connectsTo' can be configured in parallel, but the service at the source node needs to deploy only after starting the target node.  
	
	\item \emph{`deleteFrom'} connection type defines the source node type is required to be removed from the end node type.    
	
	\item \emph{`migrateTo'} connection type defines the source node type that is to be migrated to the end node type.   The 'migrateTo'  relation type cannot be defined without a 'deleteFrom' connection type to assure correctness of the business model. 
\end{enumerate}

\subsubsection{Extensibility of the Metamodel}
CloudCAMP is an opinionated framework; however, with lots of freedom. The metamodel has been designed for extensibility so that in future we can add more application node types. 
\begin{enumerate}
	
	\item If the application type is defined, e.g., SQLite needs to be added, it will go under the DBApplication branch, and all the specific attributes will be automatically inherited from the parent node e.g. DBApplication. If more attributes need to defined, the framework designer can add it under SQLite component type. 
	
	\item If the application type is not defined, such as if framework designer wants to add a stream processing engine, then StreamProcessingEngine component should be added under the parent Application node and should capture the commonalities as attributes. Then as a child of StreamProcessingEngine node type (at M1 level) specific engine, such as Apache Kafka, Storm needs to be added. The variability points specific to be the engines needs to be added as attributes. Reverse engineering can obtain the variable attributes. 
	
\end{enumerate}
Adding a new application component is time-consuming; however, it is a one-time effort, and it is reusable.

\subsection{CloudCAMP Knowledge Base Design} \label{sec:kb}  
The Knowledge Base of CloudCAMP comprises a database and the application type templates. 

\begin{figure}[htb]
	\begin{minipage}[b]{0.5\linewidth}
		\centering
		\includegraphics[width=0.99\linewidth]{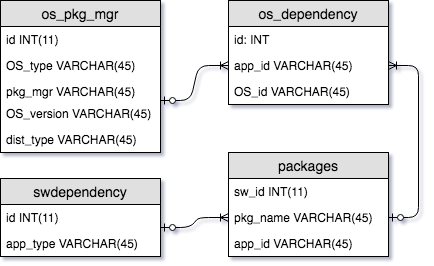}
		\caption{(a)Entity-Relation(ER) Diagram of CloudCAMP knowledge base} 
		\label{1a} 
	\end{minipage}
	\begin{minipage}[b]{0.4\linewidth}
		\centering
		\includegraphics[width=0.99\linewidth]{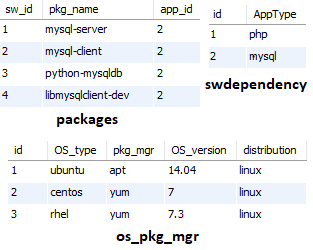}
		\caption{(b)Sample portion of KnowledgeBase Database tables} 
		\label{1b} 	
	\end{minipage}
	\caption{Knowledge Base }
	\label{fig:kb} 
\end{figure}

\subsubsection{Knowledge Base Database Design}
The ER diagram of the knowledge base database is depicted in Figure~\ref{fig:kb}(a), and it reflects the artifact sets stored in the knowledge base.
We have structured it as four tables: \emph{os\_pkg\_mgr,} \emph{os-dependency,} \emph{packages} and \emph{swdependency} to build the knowledge database. We store 1) all the operating systems, their distributions, and versions in the os\_pkg\_manager table, 2) all available application component types, e.g., PHP based web application, MySQL based DB applications, etc. are stored in swdependency table, and 3) all the software packages needed for a particular application type is found using reverse engineering and stored in the packages table.  For example, to install the scikit-learn package ({\url{ http://scikit-learn.org}}), one needs to install python, python-dev, python-pip, python-numpy,  etc. using apt-manager package, and then the scikit-learn package can be installed from the pip package manager. In a relational table called os\_dependency, we map the software packages and their versions with operating systems and their versions and store it as a key-value pair.  For instance, to install java8 on Ubuntu 16.04, we need different packages than to install java8 on windows10. We build the lookup table manually to handle these variability points.  For new application component types, the application developer needs to populate the tables with all software dependencies. The sample section of the database table structure is shown in Figure~\ref{fig:kb}(b).

\subsubsection{Knowledge Base Template Design}
\begin{figure}[htb]
	\centering
	\includegraphics[width=\linewidth]{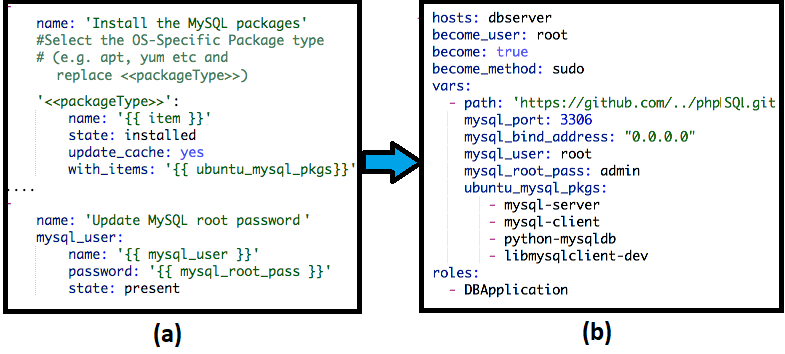}
	\caption{(a)Sample DBapplication type template and (b)Sample portion of the Auto-generated code for Deploying MySQL DB application}
	\label{Fig:kbtemplate}
\end{figure}

The knowledge base templates are designed by capturing the commonality point of the application components, and it leaves the placeholders which need to be filled up by the CloudCAMP DSML by querying the knowledge base database. One sample ansible-specific template is shown in Figure.~\ref{Fig:kbtemplate}(a). The algorithms to fill the template from the user-defined specifications are described in Algorithm~\ref{generationlogic} and ~\ref{migrationlogic}. The generated filled template is shown in  Figure.~\ref{Fig:kbtemplate}(b). Different templates are designed to serve specific application components with the different configurations. 

\subsubsection{Extensibility of the Knowledge Base}
The knowledge base is extensible by design. Addition of new
application components requires the design of new templates (at least by part) by reverse engineering the software stack. The commonalities and variabilities need to be identified, and according to that, the template needs to be designed. The software dependencies for the application components need to be identified, and this information should be inserted in the knowledge base database tables: \emph{os\_pkg\_mgr,} \emph{os-dependency,} \emph{packages} and \emph{swdependency}. Similarly, for the new application components, the framework designer should insert and manipulate the records in these tables correctly.

\subsection{Generative Capabilities of CloudCAMP DSML} 
\label{sec:meta2}  

CloudCAMP DSML provides generative capabilities for an IAC solution by interpreting the instances of models for which it incorporates a built-in knowledge base.  The CloudCAMP DSML in WebGME is built using JavaScript, NodeJS, and a MySQL database.  

As an example, we will walk through the specifications needed to be captured for the WebApplication and DBApplication component types. As shown in the M0 level of Figure~\ref{Fig:mof}, the HTTP servers (e.g., Apache web server) for the webEngines are captured in WebApplication component type, and that is related to the node template for a WebApplication.  The development languages and frameworks (Node.js, PHP, Django, etc.) of the webApplication is taken as attributes in the software property as depicted in the M1 level of Figure~\ref{Fig:mof}, which is derived from Application type of M2 level, and our modeling tools metamodel is shown in M3 level.  Similarly, as shown in the M0 level of Figure~\ref{Fig:mof}, the software for the database types (e.g., Relational Databases such as MySQL, PostgreSQL, or NoSQL databases such as Cassandra, MongoDB) are captured in DBApplication component type, and that is related to the node template for the Database Application.  Related features, such as the user id, password, specific binding port number of the Database application, etc. are stated as attributes, which is captured in the M1 level of the MOF.

\subsubsection{Knowledge Base for Generation of Infrastructure-as-code Solution for Deployment}

%

CloudCAMP's generative capabilities (Requirement \ref{sec:req2}) are enabled via
a WebGME plugin, which is invoked by a user after the modeling process.  It
generates and executes IAC as described in Algorithm~\ref{generationlogic}.
The VMs are spawned in the specified cloud platform based on the destination of
`HostedOn' connection [Lines 8-14].  Wherever possible, CloudCAMP will ensure
that scripts specific to provisioning run in parallel to provide faster
deployment.  Once the VMs are spawned, \emph{GenerateConfig()} queries the
knowledge base [line24-34] to populate the appModel [line17] based on the user's
specifications.  Then, the query result fills application-specific predefined
configuration templates and generates IAC, e.g., Ansible, for specific
application components [line 29-34] using template-based transformation.  A
similar approach is taken to configure the service-specific containers or to
start the cloud-specific services.

A sample of the automated SQL script used to query the knowledge base for deployment script generation is shown below:  

\lstset{basicstyle=\fontfamily{bch}\selectfont\scriptsize, keywordstyle=\color{blue},breaklines=true}

\begin{lstlisting}[language=sql]
SELECT  pkg.pkg_name FROM packages pkg, swdependency dep 
WHERE   pkg.app_id = dep.id 
AND     pkg.apptype = <language> 
AND     pkg.sw_id IN 
	( 
		SELECT app_sw_id 
		FROM   os_dependency 
		WHERE  os_id IN 
			( 
				SELECT id 
				FROM   os_pkg_mgr 
				WHERE  concat(os_type, os_version)=<os>, 
					<version>))
\end{lstlisting}

\begin{algorithm}
	\caption{Deployment Script Generation}\label{generationlogic} 
	\SetAlgoLined
	\DontPrintSemicolon
	\fontsize{11pt}{11}\selectfont
	
	${cloudModel} \gets \text{Objects to store cloud specs}$\;
	${appModel} \gets \text{Objects to store app specs}$\;
	\;
	Procedure GenerateIAC()\;
	\If {$ConectionType == \text{`HostedOn'}$}{
		${cloudType} \gets \text{the destination node of connection}$\;
		${appType} \gets \text{the source node of connection}$\;            
		\If {$cloudType == \text{`Desired Cloud Platform'}$}     {
			\While{!cloudModel.empty()} {
				Traverse the cloudModel\;
				Fill `cloudType' specific API Template\;
				Generate 'cloudType' specific script\;
				Execute script to spwan VMs\;
			}
		}
		${IPAddress(es)} \gets \text{IP Address of target machine}$\;        
		GenerateConfig(IPAddress(es),appType)\;   
		
		\If {$ConectionType == \text{`connectsTo'}$} {
			Find the source and destination application type\;
			Prepare workflow to execute destination script(s) first and source script later\;
		}              
	}    
	\;
	Procedure GenerateConfig()\;
	\KwIn{IPAddress(es) of Application Component Type}
	Create empty Tree Structure\;
	Fill `hosts' with IPAddress(es) of App Component Location\;
	\If {$appComponent == \text{`Desired Application Type'}$} {
		\While{!appModel.empty()} {
			Traverse the appModel\;
			Query dataBase for appType = `appComponent'\;
			Fill `appType' specific API Templates \;
			Create complete Tree Structure\;
		}
	}
	Wait for SSH in target machine(s)\;
	Run workflow to execute tasks in parallel\;                        
\end{algorithm}

\subsubsection{Determining the Order of Deployment and Execution}
The Enactor component, which is a NodeJS script, builds the dependency tree for the application types defined in the metamodel and feeds it to the orchestration workflow engine.  We generate scripts for automation tools (e.g., Ansible playbooks) for different component types, and these tools can in turn dispatch tasks to multiple hosts in parallel.  If there is a `connectsTo' relationship in the model, we let the dependent script complete first by defining the dependency chain [Line 18-21].  All the `HostedOn' dependent building blocks run in a linear fashion.  Thus, the Enactor remotely connects to the deployment hosts and deploys the application in proper order.  

\subsubsection{Generation of Infrastructure-as-code for Migration}\label{sec: migration}
The algorithm for generating a migration workflow (Requirement \ref{sec:req3}) is portrayed in Algorithm~\ref{migrationlogic}.  The `deleteFrom' connection type specifies from where the user wants to move the application components and attaches a `migrateTo' connection type to indicate the destination.  The migrationType (stateless or stateful) must be selected, and depending on that, CloudCAMP decides to checkpoint application state or not before terminating the old VMs/containers [Line 17-23].  The `migrateTo' relation type cannot be defined without `deleteFrom' connection type to ensure correctness of the model.   

\begin{algorithm}
	\caption{Migration Script Generation}\label{migrationlogic} 
	\SetAlgoLined
	\DontPrintSemicolon
	\fontsize{11pt}{11}\selectfont
	
	${cloudModel} \gets \text{Objects to store cloud specs}$\;
	${appModel} \gets \text{Objects to store app specs}$\;
	\;
	Procedure MigrationIAC()\;
	\If {$ConectionType == \text{` deleteFrom'}$}{
		${cloudType} \gets \text{the destination node of connection}$\;
		${appType} \gets \text{the source node of connection}$\;
		${IPAddress(es)} \gets \text{IP Address of target machine}$\;                
		\If {$cloudType == \text{`Desired Cloud Platform'}$} {        
			Generate 'cloudType' specific workflow script\;
			Execute script to terminate VMs\;
		}
	}
	\If {$ConectionType == \text{`migrateTo'}$} {
		GenerateIAC()\;
	}     
	\If {$migrationType == \textit{`stateless'}$}{    
		Execute deletion and migration scripts in parallel\;
	}
	\ElseIf{$migrationType == \textit{`stateful'}$}    {        
		Checkpoint current application state on old machine\;
		Restore checkpoint on the current machine\;
		Execute deletion and migration scripts in parallel\;
	}
	
\end{algorithm}

Although actions are taken for live migration, an application component from one VM to another depends on the application component type, which is a hard problem. For example, live migration of DBApplication needs a two-phase commit protocol, and a consensus algorithm to make it reliable. For the sake of simplicity, in the Algorithm~\ref{migrationlogic} we generalize our approach.  Our future work will consider more complicated scenarios of live migration and application consistency and availability issues. 

According to Algorithm ~\ref{migrationlogic}, it will spawn a new VM with the new operating system for the `migrateTo' destination node. For Stateful migration[line 20-23], our platform creates a manager node with a load balancer, and deploy the application on the current node. From that point of time, load balancer redirects all the new request to the current node, and it checkpoints the current state of the old node and restores it in the current node. Finally, it detaches the load balancer node. Thus, it produces the full infrastructure-as-code solution along with the related configuration files. All of these complete the Ansible layout tree structure helps to migrate application components from one node to another node.

\subsubsection{Support for Continuous Delivery} 
CloudCAMP can also handle continuous delivery and component addition/deletion, which is just a matter of updating the model with addition or removal of a component. For instance, to add a new database server, a user extends the model with a DBApplication node type and `connectsTo' relationships from the webserver to the database server.  CloudCAMP will generate IAC for the newly added component and executes it to deploy added component without hampering availability of the existing application. Since Ansible is idempotent, it always sets the same configuration in the target environment regardless of their current
state.

\subsubsection{Constraints checking for Correctness Business Models}
We also validate the business model by checking for constraint violations thereby ensuring that the models are ``correct-by-construction.''  We verify the correctness of the endpoint configurations for application component types, the relationship types, cloud-specific types, etc., and the business model as a whole before generating any infrastructure code.   Examples of some of the constraints are shown below:

\begin{itemize}
	\item  $ \forall$ Applications $\in$ WebApplication $\exists!$ WebEngine  
	\item  $ \forall$ Applications $\in$ DBApplication $\exists!$ DBEngine  
	\item  $ \forall$ Platform $\in$ Openstack $\exists!$ imageName  
	\item  $ \forall$ Applications $\in$ DataAnalyticsApp $\exists$
	processEngine   etc. 
\end{itemize}

We also verify other rule-based constraints to verify the components compatibility. For example, Amazon Kinesis delivery stream destination has to be Amazon Services (e.g., Redshift, S3), it cannot be Azure or OpenStack Services. We gather this information using reverse engineering.
Thus, we validate the business model by satisfying the constraints and notify the user if there are any discrepancies in  business model.


\section{EVALUATION of CloudCAMP PLATFORM}
\label{sec:CloudCAMPcasestudies}

This section describes results comparing the time and effort incurred in
deploying application use cases using (a) manual efforts, where the deployer
must log into each machine and type the commands to install packages and deploy
the applications, (b) manually writing scripts to deploy these applications,
and (c) using the CloudCAMP framework.   

\subsection{Case Study 1: LAMP-based Service Deployment Study}

\emph{\textbf{Use Case}}: This is a prototypical three-tier Linux, Apache,
MySQL, and PHP (LAMP)-based microservice architecture deployment similar to the
motivating example described in
Section~\ref{sec:CloudCAMPmotivation}. Figure~\ref{Fig:business_model} shows the
application topology illustrating the modeling effort in CloudCAMP.  

\begin{figure}[htb]
	\begin{minipage}[b]{0.4\linewidth}
		\centering
		\includegraphics[width=0.99\linewidth]{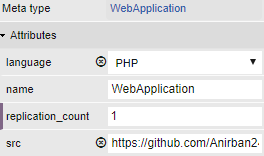}
		\caption{(a) specifications related to WebApplication type} 
		\label{1a} 
	\end{minipage}
\hspace{1cm}
	\begin{minipage}[b]{0.4\linewidth}
		\centering
		\includegraphics[width=0.9\linewidth]{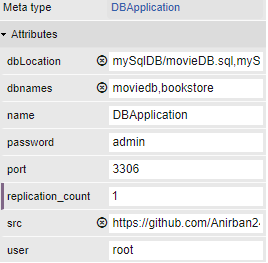}
		\caption{(b) specifications related to DBApplication type} 
		\label{1b} 	
	\end{minipage}
	\caption{ Application-specific attributes }
	\label{fig:specification} 
\end{figure}

Here, we describe the details of template-based
transformation that happens behind the scenes within CloudCAMP DSML. As stated
in  Algorithm~\ref{generationlogic},  the DSML traverses the business logic
tree of Figure~\ref{Fig:business_model}, which is defined by the deployer, and
collects all the user-defined attributes as shown in
Figure~\ref{fig:specification}.   It populates the pre-defined template for the 
specific application type with the user-defined attributes. The `mysql\_user'
and `mysql\_root\_pass' will be filled from  specifications related to
DBApplication type (Figure~\ref{fig:specification}(b)).  

The application components' software dependencies are gathered by querying the
knowledge base database. For example, to install MySQL on a Ubuntu16.04
machine, the mysql-server and mysql-client software packages are needed. So,
CloudCAMP DSML will query the knowledge base database and runs the
template-based transformation to concretize the pre-defined partial template.
The DSML copies the related configuration files in specific folders to
configure MySQL correctly. Thus, the DSML will populate the pre-defined
template file with all the details, and generate deployable Ansible-specific
deployable IAC.  
After generating all the Ansible-specific files, the CloudCAMP executes these
files in proper order to deploy the application by provisioning the cloud
infrastructure as described in Algorithm~\ref{generationlogic}. 

\subsubsection{Measure of Manual Effort:}
We conducted a small user study in a Cloud Computing course for case study 1
involving sixteen teams of three students each.  We requested users to manually
configure the files, create the handlers to specify the deployment order in the
desired host, log into each host where the application components are deployed
and manually install the packages, configure the software packages and finally
start the different components in the correct order. We have also requested
them to write the ansible script to provision the same application stack and
infrastructure.  We measured the time taken, and efforts for (a) a fully manual
effort, (b) for writing scripts in Ansible and executing these manually, and
(c) using the CloudCAMP framework to deploy the scenario.    

\emph{\textbf{Quantitative Evaluation based on a User Study:}}
The questionnaire as shown in Table~\ref{tab:survey} was created to conduct the
study.  For each question, the evaluation scale was 1--10 where one is the
easiest and ten is the hardest.  

\renewcommand{\arraystretch}{1.2}
\begin{table}[htb]	 
  \centering
  \caption{Survey Questionnaire: For Q1--Q3, rate on a scale of (1-10)} 
  \label{tab:survey}
  {\footnotesize
  \begin{tabulary}{\columnwidth}{L|L} \hline
    \multicolumn{1}{c|}{\textbf{Num}} & \multicolumn{1}{c}{\textbf{Question}} \\ \hline

    Q1 & How easy is it to deploy PHPMySQL application manually? \\ \hline

    Q2 & How easy is it to deploy PHPMySQL using DevOps tool like  Ansible? \\ \hline 

    Q3 & How easy is it to deploy PHPMySQL using CloudCAMP?\\ \hline 

    Q4 & How much time and effort did you require to deploy the application manually (in minutes)? \\ \hline 

    Q5 & How much time and effort is required in deploying the application using DevOps tool like Ansible (in minutes)? \\ \hline  

    Q6 & How much time and effort is required deploying the application using CloudCAMP (in minutes)? \\ \hline 

    Q7 & How likely are you to use the CloudCAMP platform to deploy applications in future? \\ \hline 
  \end{tabulary}
  }
\end{table}

\begin{figure}[htb]
	\begin{minipage}[b]{0.6\linewidth}
		\includegraphics[width=0.9\columnwidth]{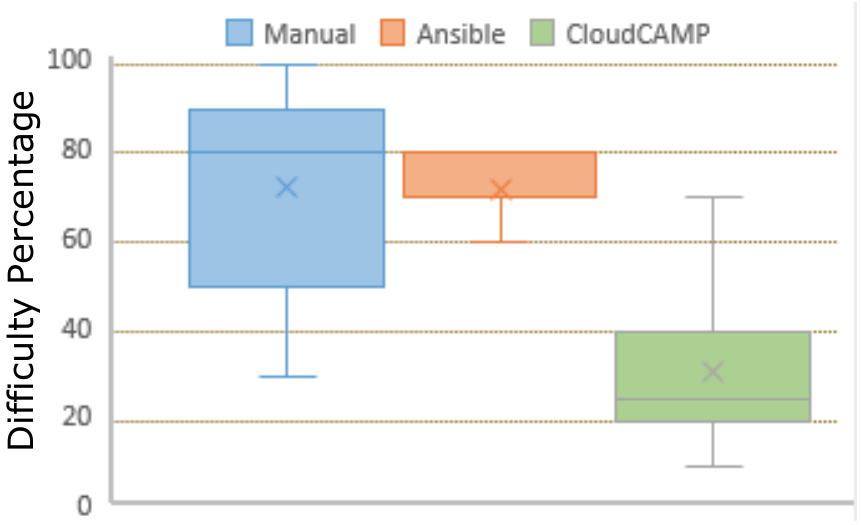}
		\caption{Comparing difficulty percentages to deploy services in \\different approaches}
		\label{Fig:difficulties}	
	\end{minipage}
	\begin{minipage}[b]{0.35\linewidth}
		\centering
		\includegraphics[width=0.75\linewidth]{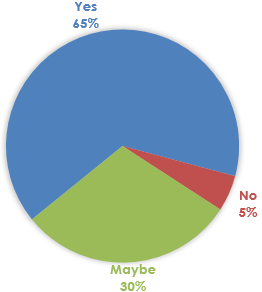}
		\caption{Likeliness of using CloudCAMP for future cloud services deployment}
		\label{Fig:future}	
	\end{minipage}
\end{figure}

\emph{\textbf{Responses to Q1, Q2, and Q3: Ease of use:}} 
As seen from Figure~\ref{Fig:difficulties}, the ``ease of use'' rating for the
CloudCAMP platform is much higher compared to manual and scripting efforts.
The median difficulty in the manual effort is rated as $72.2\%$, and median
difficulty in scripting effort is rated as $71.6\%$, while the median
difficulty rating for CloudCAMP use is $30.9\%$.   

\emph{\textbf{Responses to Q4, Q5, and Q6: Time to complete the entire deployment:}} 

The effort incurred by the user to deploy the LAMP model in the Cloud is shown
in Table~\ref{tab:survey1}, whereas using the CloudCAMP
the same topology deployment time is approximately 15-20 minutes for the first
time users.   

\begin{table}[ht]
	\centering
	\resizebox{\linewidth}{!}
	{\begin{minipage}{1.2\linewidth}
	\normalsize
	\begin{tabular}{|l|l|l|l|l|}
		\hline
		& \textbf{\begin{tabular}[c]{@{}l@{}}Deployment \\ Time(mins)\end{tabular}} & 
		\textbf{\begin{tabular}[c]{@{}l@{}}Lines to \\ Deploy\end{tabular}} & 
		\textbf{\begin{tabular}[c]{@{}l@{}}Migration \\ Time(mins)\end{tabular}} & 
		\textbf{\begin{tabular}[c]{@{}l@{}}Lines to \\ Migrate\end{tabular}} \\ \hline
		\textit{\textbf{median}} & 510 & 300 & 720 & 550 \\ \hline
		\textit{\textbf{mean $\pm$ std.dev}} & 516$\pm$244 & 315$\pm$47 & 653$\pm$231 & 553$\pm$142 \\ \hline
	\end{tabular}
\end{minipage}}
\caption{For Q5--Q6, median and mean$\pm$std.dev for deployment time, Lines of code written for deployment, migration time and Lines of code written for migration.} 
\label{tab:survey1}
\end{table}

\emph{\textbf{Response to Q7:}}As shown in Figure~\ref{Fig:future}, $65\% $ of
the respondents agreed to use CloudCAMP tool to deploy cloud applications in
the future, whereas $30\%$ are still unsure. 

\emph{\textbf{Discussion:}}
Results from our user study strengthen our belief that the CloudCAMP platform
will be a very resourceful and productive tool for business application
deployers.  We have also conducted a user study specifying to create Docker
Containers (\url{https://www.docker.com/}) and deploy the LAMP architecture
inside it using scripting tools and found very similar results.  The visual
drag and drop environment helps users to quickly deploy various scenarios of
business application topology in distributed systems.   Therefore, the benefits
of automated provisioning accrued using CloudCAMP can easily be understood.

\subsection{Case Study 2: Application Component Migration for LAMP-based Web Service}
CloudCAMP platform also supports application component migration with ease for
which we have two connection types `deleteFrom' and `migrateTo'.  As described
in Scenario~\ref{sec:req2}, suppose the user wants to migrate the database
application component from one machine to another machine, which resides on a
different OpenStack cloud platform.  This assignment was to migrate the
`stateful' MySQL database service from one node to another node, and the
students are asked to add load balancer node to make the service available all
the time.  CloudCAMP generates a new workflow structure based on the changed
user specifications as described in section~\ref{sec: migration}.  

\emph{\textbf{Responses to Q4, Q5, and Q6: Time to complete the whole migration:}}
The average time the students took to write the scripts to complete the entire
migration process is 653 minutes, with a median of 720 minutes as shown in
Table~\ref{tab:survey1}.  Whereas our rough estimates for students using the
CloudCAMP-based topology migration will be only 10-15 minutes for the first
time users. The average lines of code written using manual effort for the
migration process are 553 lines as per the survey is shown in
Table~\ref{tab:survey1}.  



\section{CONCLUSION} \label{sec:CloudCAMPconclusion}

\subsection{Summary}
This chapter presented a model-driven engineering
and generative programming approach for an automated deployment and management platform for cloud applications.  It aids the application deployer
in modeling service provisioning at a higher level of abstraction, and deploy its code without requiring significant domain expertise while requiring only
minimal modeling effort and no low-level scripting.  All the application components are the building blocks in our modeling environments and can be connected using exposed endpoints as a pipeline.  The DSML will generate
"correct-by-construction" IAC solution from the pipeline and execute the IAC to provision the application stack on the target cloud environment. Using WebGME to define the CloudCAMP framework enables us to decouple its metamodel(s) and knowledge base from the generative aspects while permitting extensibility. CloudCAMP significantly increases the productivity and efficiency of the application deployment and management team. CloudCAMP is available in open source from \url{https://doc-vu.github.io/DeploymentAutomation.}  

\subsection{Discussion}
CloudCAMP provides the flexibility to modify the application components as needed and facilitates selecting the building blocks for business requirements.
We leave the choice of application design to the application deployer. For example, one can select MongoDB and Cassandra for their backend in the development phase, deploy and do a performance test without much hassle.  As the framework matures, we can support more application components.


\section*{Acknowledgment}

{\small
	This work was supported in part by NEC Corporation, Kanagawa, Japan
	and NSF US Ignite CNS 1531079.  Any opinions, findings, and
	conclusions or recommendations expressed in this material are those of
	the author(s) and do not necessarily reflect views of NEC or NSF. 
}

\balance
\bibliography{References}{}

\begin{thebibliography}{10}
\providecommand{\url}[1]{#1}
\csname url@samestyle\endcsname
\providecommand{\newblock}{\relax}
\providecommand{\bibinfo}[2]{#2}
\providecommand{\BIBentrySTDinterwordspacing}{\spaceskip=0pt\relax}
\providecommand{\BIBentryALTinterwordstretchfactor}{4}
\providecommand{\BIBentryALTinterwordspacing}{\spaceskip=\fontdimen2\font plus
\BIBentryALTinterwordstretchfactor\fontdimen3\font minus
  \fontdimen4\font\relax}
\providecommand{\BIBforeignlanguage}[2]{{%
\expandafter\ifx\csname l@#1\endcsname\relax
\typeout{** WARNING: IEEEtran.bst: No hyphenation pattern has been}%
\typeout{** loaded for the language `#1'. Using the pattern for}%
\typeout{** the default language instead.}%
\else
\language=\csname l@#1\endcsname
\fi
#2}}
\providecommand{\BIBdecl}{\relax}
\BIBdecl

\bibitem{barve2017pads}
Y.~D. Barve, P.~Patil, A.~Bhattacharjee, and A.~Gokhale, ``Pads: Design and
  implementation of a cloud-based, immersive learning environment for
  distributed systems algorithms,'' \emph{IEEE Transactions on Emerging Topics
  in Computing}, vol.~6, no.~1, pp. 20--31, 2018.

\bibitem{bhattacharjeemde}
A.~Bhattacharjee, ``Mde-based automated provisioning and management of cloud
  applications.'' in \emph{MODELS (Satellite Events)}, 2017, pp. 480--483.

\bibitem{toscaspec}
OASIS, ``{Topology and orchestration specification for cloud applications},''
  \url{http://docs.oasis-open.org/tosca/TOSCA/v1.0/TOSCA-v1.0.pdf}, 2013, oASIS
  Standard.

\bibitem{breitenbucher2013integrated}
U.~Breitenb{\"u}cher, T.~Binz, O.~Kopp, F.~Leymann, and J.~Wettinger,
  ``Integrated cloud application provisioning: interconnecting service-centric
  and script-centric management technologies,'' in \emph{OTM Confederated
  International Conferences" On the Move to Meaningful Internet
  Systems"}.\hskip 1em plus 0.5em minus 0.4em\relax Springer, 2013, pp.
  130--148.

\bibitem{humble2011enterprises}
J.~Humble and J.~Molesky, ``Why enterprises must adopt devops to enable
  continuous delivery,'' \emph{Cutter IT Journal}, vol.~24, no.~8, p.~6, 2011.

\bibitem{carrasco2016bidimensional}
J.~Carrasco, J.~Cubo, F.~Dur{\'a}n, and E.~Pimentel, ``Bidimensional
  cross-cloud management with tosca and brooklyn,'' in \emph{Cloud Computing
  (CLOUD), 2016 IEEE 9th International Conference on}.\hskip 1em plus 0.5em
  minus 0.4em\relax IEEE, 2016.

\bibitem{eilam2011pattern}
T.~Eilam, M.~Elder, A.~V. Konstantinou, and E.~Snible, ``Pattern-based
  composite application deployment,'' in \emph{Integrated Network Management
  (IM), 2011 IFIP/IEEE International Symposium on}.\hskip 1em plus 0.5em minus
  0.4em\relax IEEE, 2011, pp. 217--224.

\bibitem{lu2013pattern}
H.~Lu, M.~Shtern, B.~Simmons, M.~Smit, and M.~Litoiu, ``Pattern-based
  deployment service for next generation clouds,'' in \emph{Services
  (SERVICES), 2013 IEEE Ninth World Congress on}.\hskip 1em plus 0.5em minus
  0.4em\relax IEEE, 2013, pp. 464--471.

\bibitem{kepes2017sepade}
K.~K{\'e}pes, U.~Breitenb{\"u}cher, and F.~Leymann, ``The sepade system:
  Packaging entire xaas layers for automatically deploying and managing
  applications,'' \emph{month}, 2017.

\bibitem{leite2014deploying}
L.~Leite, C.~E. Moreira, D.~Cordeiro, M.~A. Gerosa, and F.~Kon, ``Deploying
  large-scale service compositions on the cloud with the choreos enactment
  engine,'' in \emph{Network Computing and Applications (NCA), 2014 IEEE 13th
  International Symposium on}.\hskip 1em plus 0.5em minus 0.4em\relax IEEE,
  2014, pp. 121--128.

\bibitem{ardagna2012modaclouds}
D.~Ardagna, E.~Di~Nitto, G.~Casale, D.~Petcu, P.~Mohagheghi, S.~Mosser,
  P.~Matthews, A.~Gericke, C.~Ballagny, F.~D'Andria \emph{et~al.},
  ``Modaclouds: A model-driven approach for the design and execution of
  applications on multiple clouds,'' in \emph{Proceedings of the 4th
  International Workshop on Modeling in Software Engineering}.\hskip 1em plus
  0.5em minus 0.4em\relax IEEE Press, 2012, pp. 50--56.

\bibitem{narain2008declarative}
S.~Narain, G.~Levin, S.~Malik, and V.~Kaul, ``Declarative infrastructure
  configuration synthesis and debugging,'' \emph{Journal of Network and Systems
  Management}, vol.~16, no.~3, pp. 235--258, 2008.

\bibitem{torlak2007kodkod}
E.~Torlak and D.~Jackson, ``Kodkod: A relational model finder,'' in \emph{Tools
  and Algorithms for the Construction and Analysis of Systems}.\hskip 1em plus
  0.5em minus 0.4em\relax Springer, 2007, pp. 632--647.

\bibitem{fischer2012engage}
J.~Fischer, R.~Majumdar, and S.~Esmaeilsabzali, ``Engage: a deployment
  management system,'' in \emph{ACM SIGPLAN Notices}, vol.~47, no.~6.\hskip 1em
  plus 0.5em minus 0.4em\relax ACM, 2012, pp. 263--274.

\bibitem{di2015automatic}
R.~Di~Cosmo, A.~Eiche, J.~Mauro, S.~Zacchiroli, G.~Zavattaro, and
  J.~Zwolakowski, ``Automatic deployment of services in the cloud with aeolus
  blender,'' in \emph{Service-Oriented Computing}.\hskip 1em plus 0.5em minus
  0.4em\relax Springer, 2015, pp. 397--411.

\bibitem{di2014automated}
R.~Di~Cosmo, M.~Lienhardt, R.~Treinen, S.~Zacchiroli, J.~Zwolakowski, A.~Eiche,
  and A.~Agahi, ``Automated synthesis and deployment of cloud applications,''
  in \emph{Proceedings of the 29th ACM/IEEE international conference on
  Automated software engineering}.\hskip 1em plus 0.5em minus 0.4em\relax ACM,
  2014, pp. 211--222.

\bibitem{lascu2013planning}
T.~A. Lascu, J.~Mauro, and G.~Zavattaro, ``A planning tool supporting the
  deployment of cloud applications,'' in \emph{Tools with Artificial
  Intelligence (ICTAI), 2013 IEEE 25th International Conference on}.\hskip 1em
  plus 0.5em minus 0.4em\relax IEEE, 2013, pp. 213--220.

\bibitem{hirmer2014automatic}
P.~Hirmer, U.~Breitenb{\"u}cher, T.~Binz, F.~Leymann \emph{et~al.}, ``Automatic
  topology completion of tosca-based cloud applications.'' in
  \emph{GI-Jahrestagung}, 2014, pp. 247--258.

\bibitem{breitenbucher2014combining}
U.~Breitenbucher, T.~Binz, K.~K{\'e}pes, O.~Kopp, F.~Leymann, and J.~Wettinger,
  ``Combining declarative and imperative cloud application provisioning based
  on tosca,'' in \emph{Cloud Engineering (IC2E), 2014 IEEE International
  Conference on}.\hskip 1em plus 0.5em minus 0.4em\relax IEEE, 2014, pp.
  87--96.

\bibitem{giannakopoulos2014celar}
I.~Giannakopoulos, N.~Papailiou, C.~Mantas, I.~Konstantinou, D.~Tsoumakos, and
  N.~Koziris, ``Celar: automated application elasticity platform,'' in
  \emph{Big Data (Big Data), 2014 IEEE International Conference on}.\hskip 1em
  plus 0.5em minus 0.4em\relax IEEE, 2014, pp. 23--25.

\bibitem{bhattacharjee2018wip}
A.~Bhattacharjee, Y.~Barve, A.~Gokhale, and T.~Kuroda, ``(wip) cloudcamp:
  Automating the deployment and management of cloud services,'' in \emph{2018
  IEEE International Conference on Services Computing (SCC)}.\hskip 1em plus
  0.5em minus 0.4em\relax IEEE, 2018, pp. 237--240.

\end{thebibliography}
\bibliographystyle{IEEEtran}

\end{document}